# Understanding the effect of curvature on the magnetization reversal of three-dimensional nanohelices

J. Fullerton[1*], A. R. C. McCray[1,2], A. K. Petford-Long[1,3] and C. Phatak[1,3**]

[1] Materials Science Division, Argonne National Laboratory, Lemont, Illinois, USA
[2] Applied Physics Program, Northwestern University, Evanston, Illinois, USA
[3] Department of Materials Science and Engineering, Northwestern University, Evanston, Illinois, USA

Email: *jfullerton.anl.gov, **cd@anl.gov

**Abstract**
Comprehending the interaction between geometry and magnetism in three-dimensional (3D) nanostructures is of importance to understand the fundamental physics of domain wall (DW) formation and pinning. Here, we use focused electron beam-induced deposition to fabricate magnetic nanohelices with increasing helical curvature with height. Using electron tomography and Lorentz transmission electron microscopy, we reconstruct the 3D structure and magnetization of the nanohelices. The surface curvature, helical curvature and torsion of the nanohelices are then quantified from the tomographic reconstructions. Furthermore, by using the experimental 3D reconstructions as inputs for micromagnetic simulations we can reveal the influence of surface and helical curvature on the magnetic reversal mechanism. Hence, we can directly correlate the magnetic behavior of a 3D nanohelix to its experimental structure. These results demonstrate how control of geometry in nanohelices can be utilized in the stabilization of DWs and control of the response of the nanostructure to applied magnetic fields.

**Introduction**

With advancing nanofabrication methods, it is now possible to fabricate complex, three-dimensional (3D) magnetic nanostructures [1-13]. These structures possess intricate spin textures that can be controlled by altering the geometry of the system [4, 14, 15]. A key nanostructure in the field of 3D nanomagnetism is the nanohelix, largely studied due to its intrinsic geometric chirality and curvature. The curvature will strongly influence the spin textures that form and the resulting magnetic behavior of the system. Strong spin-chirality coupling has been observed in nanohelical and curved systems, as well as geometry driven effects on domain wall formation and motion [5, 16-18]. A helical wire is defined by its helical radius and pitch, which then describe the helical curvature and torsion, which are defined as:

$$\kappa = \frac{|r|}{r^2+c^2}, \tau = \frac{|c|}{r^2+c^2}, \quad (1)$$

respectively, where $r$ is the helical radius and $2\pi c$ is the helical pitch [19]. The technique focused electron beam-induced deposition (FEBID) allows direct fabrication of magnetic 3D nanostructures with fine geometric control [6, 20]. This allows us to directly print complex magnetic nanohelices with multiple turns and non-uniform helical curvature and torsion [1, 4, 5]. It has already been shown that different values of helical curvature and torsion can influence the formation and motion of domain walls in nanohelices [4, 21, 22]. Hence, creating a non-uniform helical curvature and torsion within the same structure could be utilized to control the locations at which domain walls form. Hence, it is important to control and quantify the 3D geometry to then allow us to influence and understand the resulting magnetic behavior as it is of interest for applications such as domain wall memories [1]. As such, 3D spintronic structures could potentially provide a route to high density, low power computing and data storage devices.

Micromagnetic simulations are a powerful tool to represent and explain experimental results [4, 5]. However, computational structures often represent idealized version of the experimental structures and can miss real geometrical variations and features, thereby not fully representing the complexity of the system. These geometrical features can be of importance as nucleation site for magnetization reversal, for domain wall pinning or cause local variations in spin textures. Therefore, to accurately predict and understand the magnetic behavior of a nanostructure we require a method to take these features into account.

In this work, we will present complementary experimental reconstructions and micromagnetic simulations of the magnetization in two nanohelices of non-uniform curvature and torsion. The helices were fabricated by FEBID and subsequently characterized by transmission electron microscopy (TEM). Through electron tomography, we reconstruct the 3D geometry of the helices. This enables us to quantify parameters that describe the nanohelix such as the Gaussian surface curvature, the helical curvature and torsion. Lorentz TEM (LTEM) was used to determine the magnetic phase shift and reconstruct projected magnetic induction maps. The reconstructed geometric structure is then directly used in micromagnetic simulations to verify the as-grown experimental states and subsequently to correlate the magnetic reversal behavior to the experimentally observed nanohelix geometry.

**Results**

Two left-handed cobalt nanohelices were fabricated by FEBID on SiO$_x$ TEM membranes. The FEBID process is shown in Figure 1(a) for a Co$_2$(CO)$_8$ precursor. During the fabrication, an accelerating voltage of 10 kV, a beam current of 33 pA and an incoming gas flux of $\Delta P = 3 \times 10^{-6}$ Torr were used. The nanohelices were fabricated using stream files allowing for their curvature to be defined using computer aided design. As such, the helical pitch of the nanohelices is controllably decreased during growth. The result is nanohelices with increasingly tighter spirals with increasing height. Figures 1(b) and (c) show bright-field TEM images of helix 1 and 2, respectively. The diameter of the nanohelices is 80 nm, but this decreases during the final turn. Helix 1 is defined to have a slower decrease of pitch than that of helix 2. To accurately quantify the geometric parameters, we performed a tomographic reconstruction of the structure of both helices. Tomographic data were collected by imaging the sample at tilt angles from −70° to +70° in steps of 2° between ±70° and ±60° and in steps of 5° between −60° and +60°, at a magnification of 8000×. The tilt series were then aligned, and the geometric structure was reconstructed using the TomoJ plugin for ImageJ and Tomviz [23, 24]. Figures 2(a) and 2(e) show the 3D reconstructions of helix 1 and 2, respectively. Both reconstructions are colored by the Gaussian surface curvature, calculated using Paraview [25]. The Gaussian curvature is different from the helical curvature, $\kappa$, defined above and instead quantifies the curving of the *surface* of a structure such as a saddle-shape, convex or concave surface [26]. Understanding Gaussian curvature is important as it affect the variation of surface normals which in turn can affect the magnetization. In both cases, the Gaussian curvature is negative (blue) on the inside of the helix and positive (red) on the outside, as expected [27]. In Fig. 2(b) and 2(f), we plot the average Gaussian curvature of the inner (blue) and outer (red) surface of the helix as a function of height for helix 1 and 2, respectively. The Gaussian curvature is mostly constant (with some variation, due to bumps on the wires) until the very top of the helices where the structures become thinner, and hence the curvature increases.

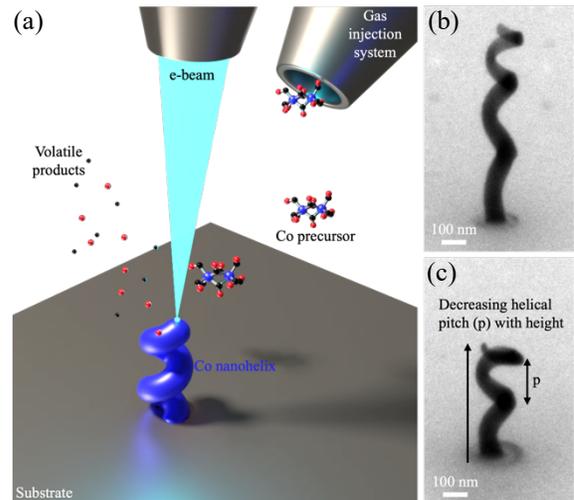

Figure 1: (a) Schematic of the FEBID process for growing magnetic nanohelices with the cobalt precursor, Co$_2$(CO)$_8$. (b, c) TEM images of fabricated nanohelices that have decreasing helical pitch with height. The structure shown in (b) will be referred to as helix 1, and the structure in (c) will be referred to as helix 2.

In Figs. 2(c) and 2(g), we quantify the helical curvature, $\kappa$, as a function of height of the nanohelix. The helical curvature increases with height due to the tighter helical spirals. For helix 1 (Fig. 2(c)) there is a slow increase of curvature at the bottom and top of the structure, with a faster increase around the mid-point. The growth of helix 2 however, begins with both a higher value of curvature and a faster increase of curvature. This rate of increase then slows down towards the top of the structure, as for helix 1. Both structures show an overall decrease in torsion from the bottom to the top, as shown in Fig. 2(d) and 2(h). This trend is expected as the torsion depends largely on the helical pitch, which decreases with height in both structures.

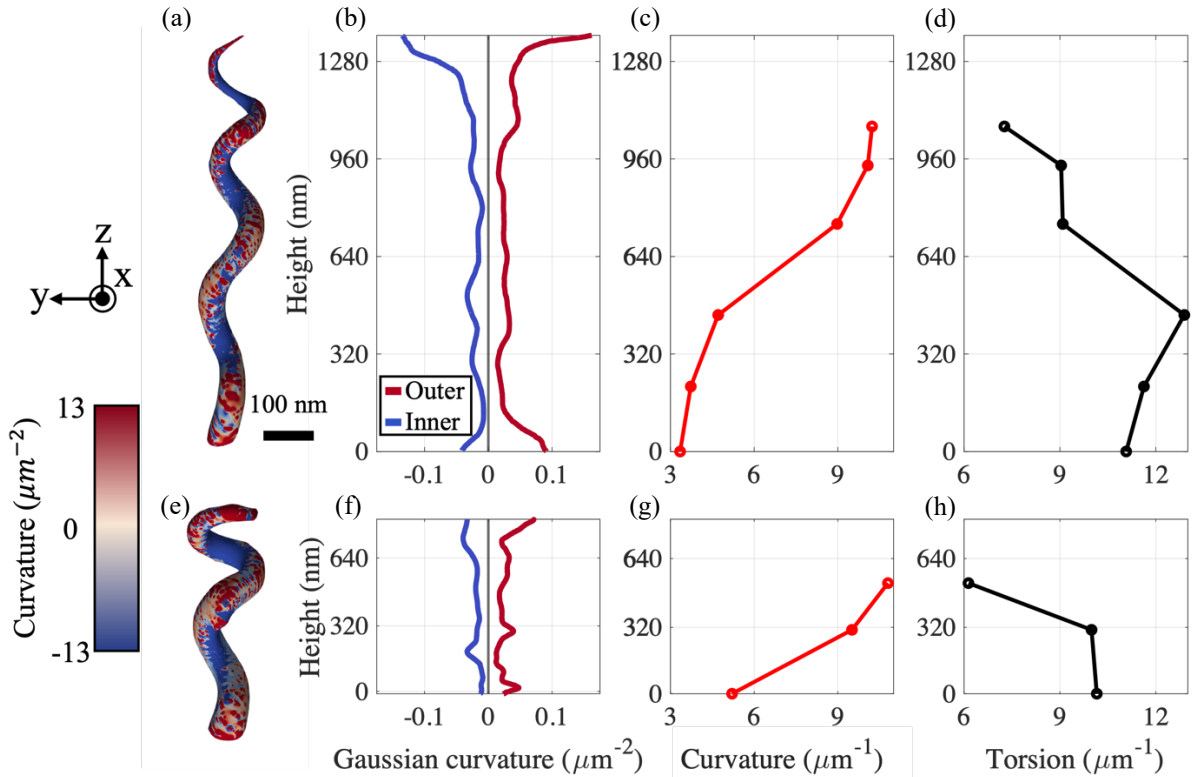

Figure 2: 3D tomographic reconstruction of the helix 1 (a) and helix 2 (e) colored by the Gaussian surface curvature (see color scale). (b, f) Averaged Gaussian surface curvature as a function of height for the inner (blue) and outer (red) surface of the helices. (c, g) Calculated helical curvature (d, h) and torsion vs. height for helix 1 and 2 respectively.

After analyzing the structure of the nanohelices, we now look to determine their magnetization through both LTEM and micromagnetic simulations, which we carried out using MuMax3 [28]. Both helices were imaged in their as-grown state, without exposure to an external magnetic field. The magnetic induction of the nanohelices was reconstructed by collecting through-focus series of images and using the transport of intensity equation method [29]. Figures 3(a) and 3(d) show colormaps of the experimentally reconstructed magnetic induction for helix 1 and helix 2 respectively. In both cases we observe an alternating pink/blue contrast along the length of the helix due to the spiraling structure. This indicates that the magnetization is likely in a single domain state where the spins follow the shape of the helix along its length.

The tomographic reconstructions, shown in Fig. 2, not only allow analysis of the geometry but can aid in understanding the magnetic behavior of the nanohelices. The reconstructed mesh that defines the surface of the helix geometry can be discretized into cubes (3 × 3 × 3 nm in this case, with edge smoothing to avoid staircase artefacts) and input into MuMax3 to allow the magnetic state of each helix to be directly simulated. A saturation magnetization of 900 kAm$^{-1}$ and an exchange stiffness of $10^{-11}$ Jm$^{-1}$ was used to be representative of FEBID cobalt [5, 12]. Figs. 3(c) and 3(f) show simulated single domain states for both nanohelices (initialized as a uniform magnetization with $M_z$ along –z for helix 1, and along +z for helix 2). The reconstructed magnetic phase images from the simulations in Figs. 3(b) and 3(e) match those of the experiments, confirming the presence of a single domain as-grown magnetization state in both helices. At the base of the simulated helices there appears to be a vortex-like magnetic configuration that is more pronounced in helix 2 (Fig. 3(f)). This is likely a consequence of the stray field produced at the ends and the fact that the structure is wider at the base. These magnetic vortices would be difficult to see in the reconstructed phase however, due to the tilt angle required when imaging.

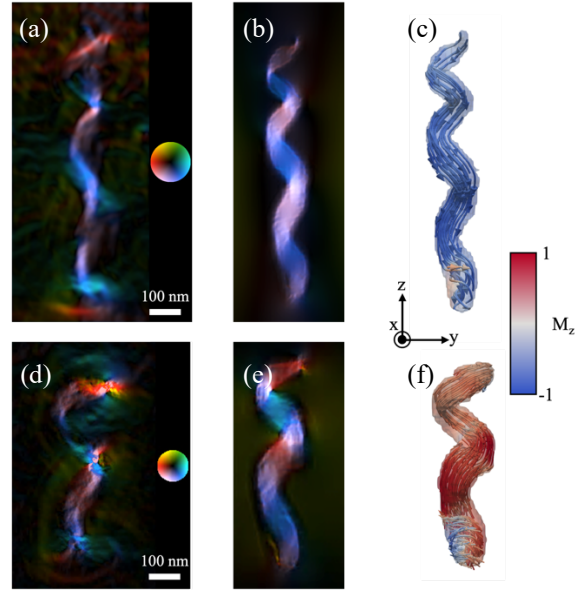

Figure 3: (a, d) Experimentally reconstructed projected magnetic induction maps. (c, f) Simulated remanent magnetic states using the geometries determined from the experimental tomographic reconstructions. (b, e) Simulated magnetic induction maps calculated from the magnetic configurations shown in (c) and (f) for helix 1 (a, b, c) and 2 (d, e, f).

In addition to verifying the as-grown magnetic states, the 3D reconstructions can allow us to further explore the magnetic behavior of the nanohelices, giving insights into the effect of varying helical curvature and torsion on the magnetization reversal behavior. In Figure 4, we plot the magnetization reversal process in the presence of an axial magnetic field, applied opposite to the initial uniform magnetic states. The field was applied at an angle that is 1° off the z-axis to break the symmetry of the simulation (to avoid computational artefacts) and was increased in steps of 0.5 mT. Fig. 4(a) shows this process for helix 1, starting from a single domain state with $M_z$ along −z. At an applied field of 25 mT, the start of the magnetization reversal process is nucleated at the bottom of the helix. Here, the magnetization becomes more vortex-like as the $+M_z$-component of magnetization is increased. At 26 mT, a small growth of the area with a positive $M_z$ is seen (shown in red). At 26.5 mT, the bottom two turns of helix 1 switch rapidly to a state with a positive value of $M_z$. This creates a vortex domain wall at the top of the helix where the curvature is higher (shown in more detail in Fig. 4(b)). The exact location of this domain wall coincides with an area of increased Gaussian surface curvature, which likely acts as a pinning location. The domain wall remains until an applied field of 38 mT at which field the magnetization in the helix has been fully reversed. The process observed here shows that variations in surface curvature can help nucleate and stabilize domain walls. Hence, the surface curvature could be used to control the location and propagation of domain walls.

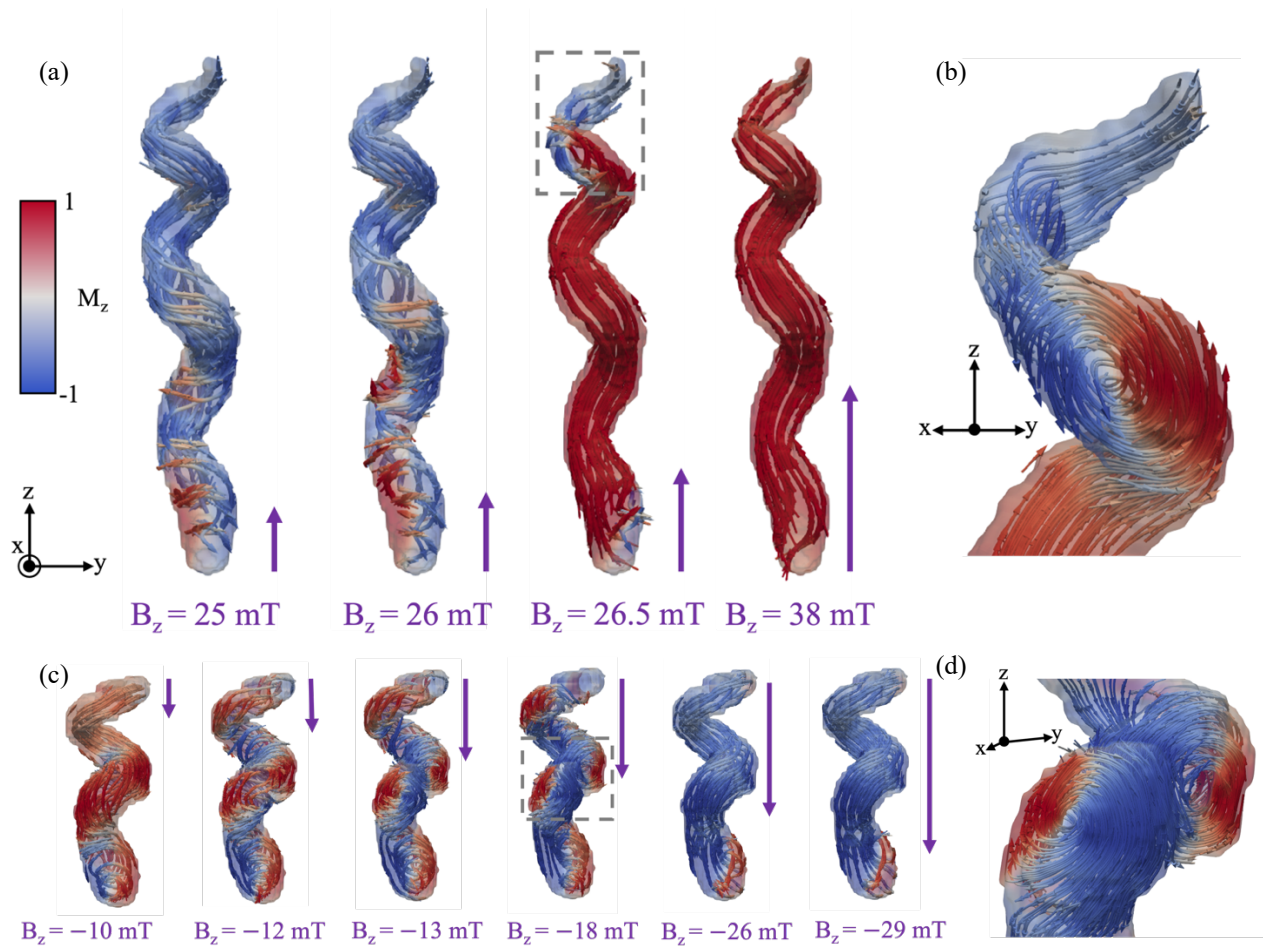

Figure 4: Simulated magnetization reversal using the tomographic reconstructions of helix 1 and 2, respectively (a, c). (b) Close-up view of the region enclosed in the gray box in (a), showing the vortex domain wall created during reversal in helix 1 at 26.5 mT. (d) Close-up view of the region enclosed in the gray box in (b), showing the double vortex texture formed during reversal in helix 2 at −18 mT.

In Fig. 4(c), we explore the reversal process for helix 2 by applying an increasing field along the −z direction. As with helix 1, the reversal process in helix 2 is nucleated from the base of the structure. This is because it is wider than the top, and hence more suitable for the vortex-like textures that form.

With increasing negative field, the vortex at the base of the helix propagates through the structure as a domain wall leading to reversal of the magnetization. Linked vortex textures form during this process, allowing the growth of regions with negative $M_z$ regions and shrinkage of regions with positive $M_z$. Fig. 4(d) shows a close-up view of the pair of vortices that form in the center of the structure at an applied field of −18 mT (location shown by the gray box in Fig. 4(c)). Here, we observe two left-handed vortices where the central region with negative $M_z$ region grows with applied field. At an applied field of −26 mT, the field fully reverses the magnetization relative to the initial state except at the base of the helix where the vortex remains until higher fields. The increased curvature and curvature gradient of helix 2, with respect to that of helix 1, allows the stabilization of complex non-linear spin textures during reversal. This is largely due to the wire becoming more transverse to the applied field with increased helical curvature. Most of the magnetization in helix 1 is reversed over a range of 1.5 mT, with a further 11.5 mT required to expel the vortex domain wall pinned at the top. The reversal process for helix 2, however, is a more gradual process over a range of 16 mT. However, for both helices we observe that regions of higher curvature and lower torsion promote the formation and stability of non-linear spin textures such as vortices and domain walls.

The experimentally fabricated nanohelices discussed above display a well-defined trend of increasing curvature with height. Through micromagnetic simulations, we can correlate the magnetic behavior with the geometric curvature. However, the trend of torsion as a function of height is less clear. Therefore, we now discuss the behavior of three simulated helices (shown in Figure 5) each with a fixed curvature but with torsion that increases linearly with height. As previously discussed, the geometry of a helical wire depends on the helical radius and pitch, which also define the helical curvature and torsion. Hence, we can parameterize the equation of a helix to depend only on the curvature and torsion:

$$X = \frac{|\kappa|}{\kappa^2 + \tau^2}\cos(H), Y = \frac{|\kappa|}{\kappa^2 + \tau^2}\sin(H),$$
$$Z = 2\pi\frac{|\tau|}{\kappa^2+\tau^2}H, (2)$$

where X, Y and Z are the coordinates of the helical wire (see Fig. 5), and H is the height. Using these equations, we defined three helical structures with values of the torsion, $\tau$, increasing linearly between 0.05 and 25 $\mu m^{-1}$ and with constant values of the helical curvature, $\kappa$, = 20, 10 and 5 $\mu m^{-1}$. Figs. 5(a), (c) and (e) show the relaxed, remanent magnetic state for each helix initialized with a uniform value of $+M_x$ (left) or $+M_y$ (right). An initial magnetization transverse to the long axis of the helix was used to encourage the formation of domain walls within the structure. For each value of curvature, the domain walls that form are asymmetric transverse domain walls and typically form for values of $\tau = 8$–$16$ $\mu m^{-1}$, showing that domain walls are more stable with intermediate vales of torsion. Only one domain wall is observed outside this range of torsion values, and this forms for $\tau = 5$ $\mu m^{-1}$ and $\kappa = 5$ $\mu m^{-1}$ (Fig. 5(c) left), with the stability being increased by having a curvature and torsion of a similar magnitude. In comparison to the simulations using the nanohelix shapes obtained experimentally from the tomographic reconstructions, we see that domain walls are stabilized at areas of similar magnitude of torsion and curvature (*i.e.,* further up the helices). For all three nanohelices shown in Fig. 5, no domain walls are formed at areas of high torsion. Hence, the trend of decreasing torsion seen towards the top of both experimental helices may also assist in stabilizing the domain walls that form.

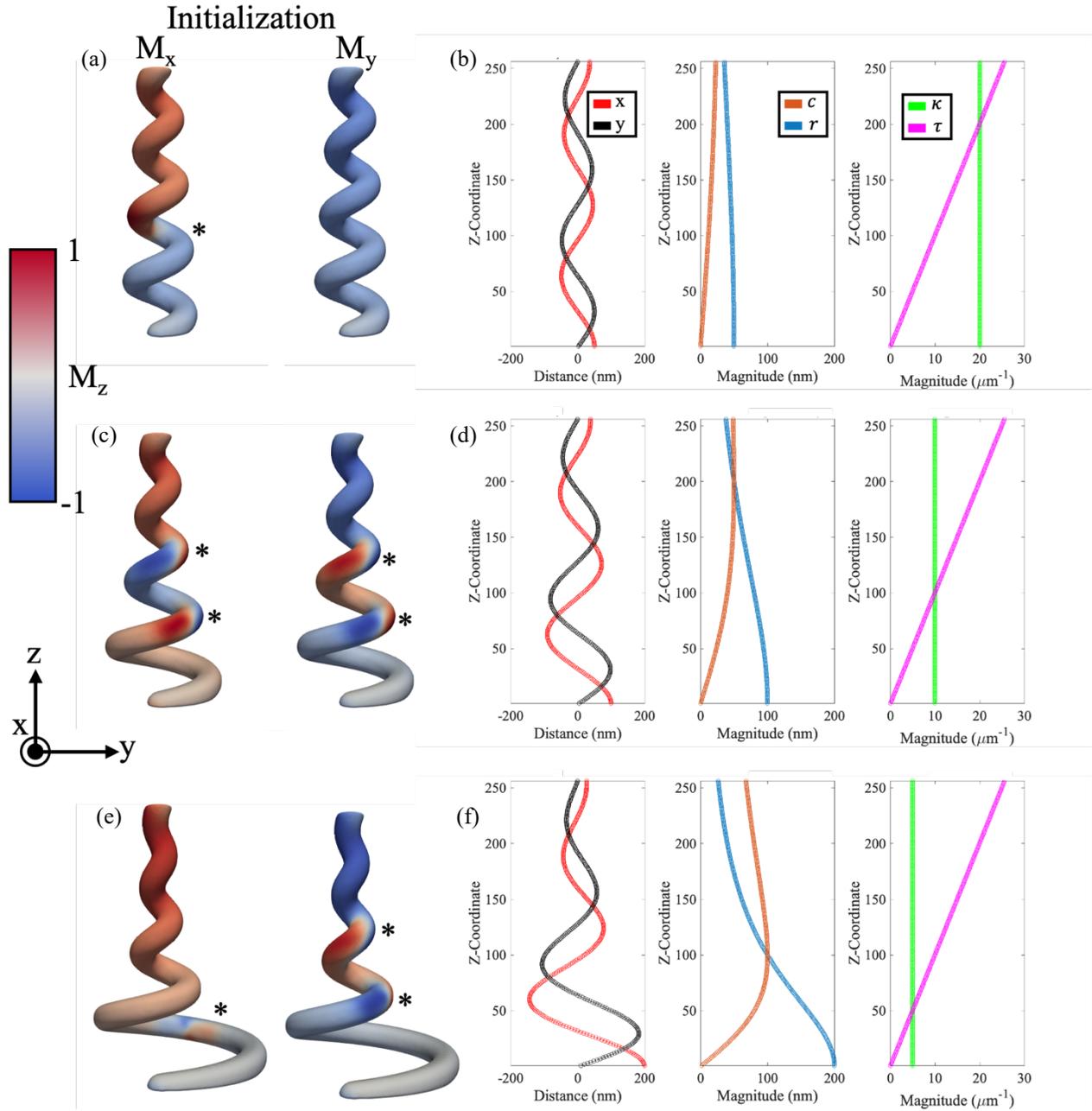

Figure 5: Simulated remanent magnetic states after initial magnetization configurations with uniform values of $M_x$ (left) or $M_y$ (right), for $\kappa = 20$ (a), 10 (c) and 5 (e) $\mu m^{-1}$. The nanohelices are colored according to the $M_z$ component, and domain walls are marked with *. Graphs of the x- and y-coordinate of the helical nanowire (left), helical pitch and radius (middle), and helical curvature and torsion (right) as a function of z-coordinate for $\kappa = 20$ (b), 10 (d) and 5 (f) $\mu m^{-1}$.

**Conclusion**

In summary, we have utilized FEBID to create nanohelices with curvature and torsion that vary with height. We have demonstrated the ability to controllably create nanostructures of complex geometry, and by using electron tomographic reconstruction of the nanohelices in 3D, we can obtain quantitative information about the surface curvature, helical curvature and torsion as a function of height. We then used the 3D reconstructions as inputs for the shape of the nanohelices in micromagnetic simulations, whereby we were able to verify the single domain states observed in the as-grown magnetic phase reconstructions of the nanohelices. Our simulations also enabled us to explore the magnetization reversal behavior of the two nanohelices, and to link the spin textures that form directly to the experimentally fabricated geometry. Variations in surface curvature can provide pinning sites for domain walls, while areas of higher helical curvature allow the formation of more non-linear spin textures during the reversal process. Further simulations on computationally defined structures revealed that intermediate values of torsion (of comparable magnitude to the helical curvature) are required for initialization of domain walls, which was also reflected in the experimental structures.

The results presented here aid in understanding how 3D geometric effects could be integrated into domain wall-based devices. Further experiments could include the fabrication of nanohelices of controlled varying curvature and torsion in order to precisely create and move domain walls. We would then look to initialize and experimentally reconstruct non-linear 3D magnetic textures through vector field tomography [30, 31].

**Methods**

*Nanofabrication.* The two cobalt nanohelices were fabricated using focused electron beam-induced deposition in an FEI Nova dual beam FIB/SEM. The precursor gas used was $Co_2(CO)_8$ with a working pressure of $3 \times 10^{-6}$ Torr, accelerating voltage of 10 kV and a beam current of 33 pA. The structures were grown using stream files that contained a series of x and y pixel coordinates and dwell times for each point. The nanohelices were grown directly onto $SiO_x$ TEM grids before being transferred to the TEM with minimal exposure to air and external magnetic fields.

*Transmission electron microscopy.* The electron tomography and magnetic phase reconstruction were performed in a JEOL 2100F LTEM instrument operating at 200 kV with an imaging $C_S$ corrector. The samples sit in a low field environment to avoid affecting their as-grown magnetic state.

Tomographic data were collected by imaging the sample at tilt angles from $-70°$ to $+70°$ in steps of $2°$ between $\pm 70°$ and $\pm 60°$ and in steps of $5°$ between $-60°$ and $+60°$, at a magnification of 8000x. The images were then aligned using cross-correlation using the TomoJ plugin in ImageJ. This aligned tilt series was used to create the 3D reconstructions using TomoViz. The surface Gaussian curvature was calculated using the curvature filter in Paraview. The data was smoothed to remove high curvature artefacts, but not overly smoothed such that the structure was altered. The 3D reconstruction meshes were discretized using the Voxelize function in PyVista. A model with voxels of size 3x3x3 nm was created and inputted in to MuMax3 to simulate the magnetization of the nanohelices.

The reconstruction of the magnetic phase of the nanohelices was performed by taking a series through-focus bright field images from $-640$ to $+640$ $\mu m$ (in 15 steps). A second through-focus series was recorded with the samples flipped $180°$ to be able to separate the magnetic and electrostatic phase shift. The retrieval of the magnetic phase was performed using the transport of intensity equation method using the open source PyLorentz code developed in our group. PyLorentz was also used for simulating the magnetic phase from the output of the MuMax3 simulations of the nanohelices.

**Acknowledgments**

This work was funded by the US Department of Energy, Office of Science, Office of Basic Energy Sciences, Materials Science and Engineering Division. Work performed at the Center for Nanoscale Materials, a U.S. Department of Energy Office of Science User Facility, was supported by the U.S. DOE, Office of Basic Energy Sciences, under contract no. DE-AC02-06CH11357.